\documentclass[reprint,superscriptaddress,aps,amsmath]{revtex4-2}

\usepackage{showyourwork}
\usepackage{graphicx}
\usepackage{mhchem}
\usepackage[separate-uncertainty=true]{siunitx}
\begin{document}

% Title
\title{Advice on describing Bayesian analysis of neutron and X-ray reflectometry}
\thanks{This work was developed as a part of the Open Reflectometry Standards Organisation Workshop in 2022 with contributions from all authors as part of a round-table discussion.}

% Author list
\author{Andrew R. McCluskey}
  \email{andrew.mccluskey@ess.eu}
  \affiliation{European Spallation Source ERIC, P.O. Box 176, SE-221 00, Lund, SE}
\author{Andrew J. Caruana}
  \affiliation{ISIS-Neutron and Muon Source, Rutherford Appleton Laboratory, Didcot, Oxon OX11 0QX, GB}
\author{Christy J. Kinane}
  \affiliation{ISIS-Neutron and Muon Source, Rutherford Appleton Laboratory, Didcot, Oxon OX11 0QX, GB}
\author{Alexander J. Armstrong}
  \affiliation{ISIS-Neutron and Muon Source, Rutherford Appleton Laboratory, Didcot, Oxon OX11 0QX, GB}
\author{Thomas Arnold}
  \affiliation{European Spallation Source ERIC, P.O. Box 176, SE-221 00, Lund, SE}
\author{Joshaniel F. K. Cooper}
  \affiliation{ISIS-Neutron and Muon Source, Rutherford Appleton Laboratory, Didcot, Oxon OX11 0QX, GB}
\author{David L. Cortie}
  \affiliation{Australian Nuclear Science and Technology Organisation, Lucas Heights, New South Wales, AU}  
\author{Arwel V. Hughes}
  \affiliation{ISIS-Neutron and Muon Source, Rutherford Appleton Laboratory, Didcot, Oxon OX11 0QX, GB}
\author{Jean-Fran\c{c}ois Moulin}
  \affiliation{German Engineering Material Science at Heinz Maier-Leibnitz Zentrum, Helmholtz-Zentrum Hereon, Lichtenbergstraße 1, 85748 Garching, DE}
\author{Andrew R. J. Nelson}
  \affiliation{Australian Nuclear Science and Technology Organisation, Lucas Heights, New South Wales, AU}  
\author{Wojciech Potrzebowski}
  \affiliation{European Spallation Source ERIC, P.O. Box 176, SE-221 00, Lund, SE}
\author{Vladimir Starostin}
  \affiliation{Institute of Applied Physics, University of T\"{u}bingen, Auf der Morgenstelle 10, 72076 T\"{u}bingen, DE}
  
\collaboration{Open Reflectometry Standards Organisation}

% Abstract with filler text
\begin{abstract}
  Driven by the availability of modern software and hardware, Bayesian analysis is becoming more popular in neutron and X-ray reflectometry analysis.
  The understandability and replicability of these analyses may be harmed by inconsistencies in how the probability distributions central to Bayesian methods are represented in the literature. 
  Herein, we provide advice on how to report the results of Bayesian analysis as applied to neutron and X-ray reflectometry. 
  This includes the clear reporting of initial starting conditions, the prior probabilities, and results of any analysis, and the posterior probabilities that are the Bayesian equivalent of the error bar, to enable replicability and improve understanding. 
  We believe that this advice, grounded in our experience working in the field, will enable greater analytical reproducibility among the reflectometry community, as well as improve the quality and usability of results. 
\end{abstract}

\maketitle

% Main body with filler text
\section{Introduction}\label{sec:intro}

Neutron and X-ray reflectometry are powerful tools to probe the interfacial structure of materials~\cite{lovell_analysis_1999}.
However, as a result of the ``phase problem'', the analysis of these techniques is ill-posed in nature, as there are multiple possible solutions~\cite{majkrzak_exact_1995}.
This has led to the use of Bayesian analysis, where some prior understanding of the system is used to aid our understanding of some reflectivity profile~\cite{sivia_analysis_1991,geoghegan_experimental_1996,sivia_bayesian_1998}.
Recently, developments in the availability of computer software for reflectometry analysis that include Bayesian functionality, such as Refl1d, refnx, anaklasis, and RasCAL~\cite{kienzle_refl1d_2021,nelson_refnx_2019,koutsioubas_anaklasis_2021,hughes_rascal_2019}, which implement sampling methods from bumps, emcee, and dynesty~\cite{kienzle_bumps_2021,foremanmackey_emcee_2019,speagle_dynesty_2020}, have led to an increase in the utilisation of Bayesian methods by the reflectometry community~\cite{mccluskey_bayesian_2019,mccluskey_general_2020}.

This work will focus on the best practice for reporting the results from Bayesian and sampling-based analysis of neutron and X-ray reflectivity data. 
This work will not introduce Bayesian or sampling methods for neutron and X-ray reflectometry analysis. For those unfamiliar with these techniques, we suggest the work of Sivia and co-workers~\cite{sivia_bayesian_1998,sivia_data_2006} and more recent work focusing on reflectometry analysis~\cite{hughes_physical_2019,mccluskey_general_2020,nelson_refnx_2019,aboljadayel_determining_2021}. 
We hope that this work will inform best practices in data sharing from reflectometry analysis and inspire software developers to enable these to be accessed easily by the user. 

Reflectometry analysis can be described, in the most simplistic terms, as a comparison and refinement of a model based on some parameters, $\mathbf{x}$, to reproduce some reflectivity data set, $\mathbf{D}$. 
This refinement process involves comparing the model to the data and calculating some goodness-of-fit value or likelihood, $p(\mathbf{D} | \mathbf{x})$, and modifying the model to optimize the goodness-of-fit or maximise the likelihood.
A commonly used goodness-of-fit parameter is the $\chi^2$ parameter which is found as~\cite{nelson_refnx_2019}, 
\begin{equation}
    \chi^2 = \sum_{q=q_{\text{min}}}^{q_{\text{max}}}{\bigg{[\frac{(R(q) - R{(q)}_{\text{m}})}{\sigma_R(q)}\bigg]}^2}, 
    \label{equ:chi2}
\end{equation}
where $R(q)$ and $R(q)_m$ are the measured and modelled reflectivity at a given $q$, respectively, while $\sigma_{R}(q)$ is the uncertainty associated with the measured reflectivity at each $q$.
Here, $q = (4\pi / \lambda)\sin{\theta}$ is the measured momentum transfer, where $\theta$ is half the scattering angle and $\lambda$ is the wavelength of the incident radiation.
Under an assumption of normally distributed residuals $R(q) - R{(q)}_{\text{m}} \sim \mathcal{N}(0, \sigma_R(q))$, the likelihood is related to the $\chi^2$ variable in the following way:
\begin{equation}
    \ln[p(\mathbf{D} | \mathbf{x})] = -\frac{1}{2} \bigg(\chi^2 + \sum_{q=q_{\text{min}}}^{q_{\text{max}}}\ln{\big[2\pi\sigma_R{(q)}^2\big]}\bigg).
    \label{equ:likelihood}
\end{equation}
The input for this refinement process is the model and some initial parameter values, which may be an absolute value or some parameter range, depending on the refinement algorithm.
The output is a set of values for $\mathbf{x}$, potentially with associated error bars --- when these are present they typically describe a standard deviation from the mean of a Gaussian probability distribution. 
While for Bayesian sampling processes, the input is a probability distribution for each parameter, known as the prior.
The sampling process gives a probability distribution, the posterior, that defines the relative likelihood of different values of each parameter, from this we can report statistical measures, e.g.\ mode/mean/median.
This process implicitly assumes that the data is completely reduced, accounting for all experimental parameters, uncertainties are accurately described and the model can accurately describe the data. 

The input required depends on a minimisation algorithm being used, with some algorithms requiring a single starting guess (such as traditional Newtonian methods) and others taking a range of potential values (more common in stochastic approaches, like differential evolution). 
The nature of these inputs defines the results of the analysis, therefore it is of the utmost importance that these are shared as part of a publication describing the work. 
Furthermore, the minimisation is often performed with bounds in place, defining that the parameter values will lie within a given range. 
This range can be thought of as having a prior probability distribution, $p(\mathbf{x})$, where values of $\mathbf{x}$ outside of this range have a probability of \num{0}. 
Even when a non-Bayesian approach is used in the analysis (i.e. Bayes theorem is not utilised), the result where bounds are set would be analogous to a Bayesian analysis with a uniform prior probability. 

The optimised parameters from the minimisation algorithm, which depend on the particular algorithm, often include some statistical uncertainty.
This uncertainty comes from an assumption of normally distributed parameters~\cite{bevington_data_2002}, however, Bayesian sampling approaches make no assumption of an underlying statistical distribution.
How these statistical uncertainties are found is beyond the scope of this work, but it is important to acknowledge that this uncertainty typically assumes that the probability distribution of the parameter is Gaussian in nature. 
This probability distribution is either the partial likelihood or posterior, the latter when some prior is included and Bayes theorem is applied. The posterior describes our understanding of the parameter values given the data that was measured.
When Bayesian modelling is used and the prior is included, the posterior probability is found as, 
\begin{equation}
  p(\mathbf{x} | \mathbf{D}) \propto p(\mathbf{D} | \mathbf{x}) p(\mathbf{x}).
  \label{equ:bayes}
\end{equation}
Therefore, when Bayesian modelling is performed, the priors and likelihood are of fundamental importance to the results that are obtained (the posterior), and any scientific conclusions that are drawn. 
We note that Equation~\ref{equ:bayes} omits the normalisation term, the Bayesian evidence [$p(\mathbf{D})$], which is discussed in detail elsewhere~\cite{sivia_bayesian_1998,mccluskey_general_2020} and can be omitted when model comparison is not being performed. 

The use of Bayesian inference can be valuable in the interpretation of reflectivity data, however, inconsistency in the description of the process will result in an analysis that cannot be reproduced or easily understood. 
This can range from not reporting the priors applied to each parameter (e.g.\ the lower/upper limits for a uniform distribution that applies box bounds), to failing to share the complete sampling chain of a Markov Chain Monte Carlo sampling, or details of any autocorrelation analysis (the last of which, the authors of this work admit to being guilty of~\cite{mccluskey_bayesian_2019}). 
In this letter, we outline the best practice for reporting the results of Bayesian analysis for neutron and X-ray reflectometry, we hope that this work will engage others to carefully consider how they report this information. 
Furthermore, uptake of the approaches discussed herein will lead to greater clarity about the models and assumptions used in, and the reproducibility of, our analyses.

\section{Prior}\label{sec:prior}

The most common probability distributions that are used for a prior are uniform between a lower and upper bound or a half-closed interval, where only a lower or upper bound is defined.
The use of a bounded parameter along with some traditional $\chi^2$-minimisation method and a parameter with a uniform prior and a Bayesian maximum \emph{a-posteriori} approach will lead to the same result. 
For priors that are uniform is it important that the upper and lower bounds are reported, this can be achieved with a simple table  (see Table~\ref{tab:bounds} for an example) to be included in the article or supplementary information. 
Note, that this table also gives information on ``constrained'' values, where in some analysis parameters were not allowed to vary, these constrained values can have a significant impact on the result of any analysis and therefore must also be given.
\begin{table}
    \caption{An example of the presenting uniform priors in a tabular format. Reproduced from~\cite{mccluskey_general_2020}, where each parameter was either constrained to a given value or sampled within the prior range.}\label{tab:bounds}
    \begin{tabular}{c | c | c}
        Parameter & Constrained Value & Prior Range \\ 
        \hline
        $d_h$/\si{\angstrom} & $10.0$ & $[8.0, 16.0)$\\
        $V_h$/\si{\angstrom^3} & $339.5$ & $[300.0, 380.0)$\\
        $d_t$/\si{\angstrom} & $21.0$ & $[10.0, 26.0)$\\
        $\phi_t$ & $1.0$ & $[0.5, 1.0]$\\
        $V_t$/\si{\angstrom^3} & $850.4$ & $[800.0, 1000.0)$\\
        $\sigma$/\si{\angstrom} & $2.9$ & $[2.9, \infty)$\\
    \end{tabular}
\end{table}

Currently, the use of non-uniform, informative priors is less common in reflectometry analysis.
However, the increasing popularity of Bayesian methods and interest in using complementary methods for analysis means that these are likely to become more popular in the coming years. 
Here we will define two potential types of informative prior probabilities, those that can be described with some mathematical function and those that cannot, for example arising from the application of a sampling-based analysis of a complementary technique. 

\begin{figure}
  \script{prior.py}
  \includegraphics[width=\columnwidth]{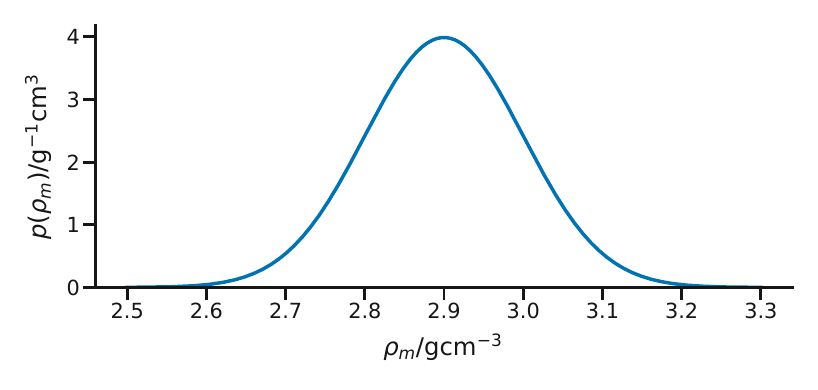}
  \caption{
    A prior probability distribution for \ce{Si3N4} with a density of $\rho_m=\SI{2.9\pm0.1}{\gram\per\cubic\centi\meter}$. 
  }\label{fig:prior}
\end{figure}

When a prior probability can be described with a mathematical function, this should be done by providing this function in the clearest possible language. 
For example, if the prior is taken from a single complementary measurement that is defined as a value with some uncertainty, which represents a normal distribution with a mean and standard deviation, this information should be provided. 
This is shown in Figure~\ref{fig:prior} for the density of silicon nitride (\ce{Si3N4}) produced by atomic layer deposition~\cite{knoops_atomic_2015} which is used to inform the value for a scattering length density for some layer of the material.
Such a prior probability distribution could be described graphically, Figure~\ref{fig:prior}, or in prose as being ``normally distributed with a mean of \SI{2.9}{\gram\per\cubic\centi\meter} and a standard deviation of \SI{0.1}{\gram\per\cubic\centi\meter}'', mathematically as, 
\begin{equation}
  p(\rho_m) = \frac{1}{\sigma\sqrt{2\pi}}\exp\Bigg[-\frac{1}{2}\bigg{({}\frac{\rho_m-\mu}{\sigma}\bigg)}^2\Bigg]
\end{equation}
where, $\mu=\SI{2.9}{\gram\per\cubic\centi\meter}$ and $\sigma=\SI{0.1}{\gram\per\cubic\centi\meter}$, or more concisely $p(\rho_m) \sim \mathcal{N}(\mu=\SI{2.9}{\gram\per\cubic\centi\meter}, \sigma=\SI{0.1}{\gram\per\cubic\centi\meter})$. 
The same descriptive approach could be taken for any common statistical distribution, including log-normal or truncated normal distributions. 

\begin{figure}
  \script{samples.py}
  \includegraphics[width=\columnwidth]{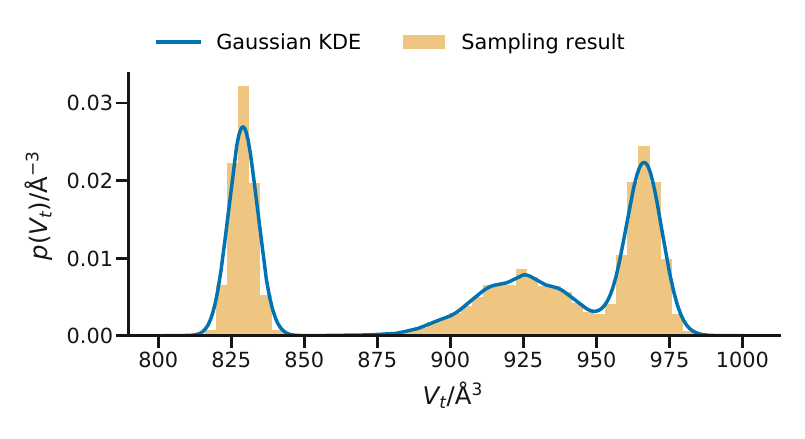}
  \caption{
    A hypothetical prior probability distribution for a DPPC lipid that could arise from a molecular dynamics simulation (orange histogram) and a Gaussian kernel density estimation for the probability distribution using a bandwidth factor of \num{0.05} (blue line). 
  }\label{fig:samples}
\end{figure}

It is possible that the prior distribution cannot be described with a simple mathematical function, with multi-modal priors being an example, if it is a multimodal model result from some other sampling method, then the chain from the prior sampling should be given. 
The chain is all of the samples investigated in the sampling and should be shared, although in the case that this chain is very large a subsampled object may be shared, in which case, the autocorrelation analysis performed should be described (see Section~\ref{sec:posterior} for a more complete discussion of this).
To use such a prior probability in Bayesian analysis, some functional description of the prior must be generated, most commonly this will be some kernel density estimation, when this is used it is also necessary to state the structure of the kernel being used. 
An example of this is shown in Figure~\ref{fig:samples}, where the prior probability for the volume of a phospholipid tail could be found from molecular dynamics simulation where there are three common conformers that the lipid is likely to have.

\section{Likelihood}\label{sec:likelihood}

Bayes theorem (given in Equation~\ref{equ:bayes}) consists of the product of the prior and the likelihood. 
The former describes our current understanding of the parameters before we conduct some experiments, while the latter describes how well the data is described by the model parameters.
Although Equation~\ref{equ:likelihood} is a common approach to quantify how well the data is described, it assumes a normally distributed uncertainty for the measured reflectivity value.
While a normally distributed uncertainty is the most common, it may not be accurate in all circumstances, for example, when low numbers of counts are present, a Poisson uncertainty may offer a more accurate description, in which case the likelihood function could be changed to that which reflect multi-dimensional Poisson distribution~\cite{lass_multinomial_2021}.
Additionally, the likelihood function may be modified by weighting the data at high-$q$, such as by replacing $R(q)$ in Equation~\ref{equ:chi2} with $\log{[R(q)]}$ or $R(q)q^4$.
These aspects make it very important to state explicitly how the likelihood for a given model and data is calculated (giving the analysis package used, version number, and if not the default the likelihood option).

\section{Posterior}\label{sec:posterior}

Bayesian analysis methods typically involve using some sampling process, such as Markov chain Monte Carlo, to estimate the posterior probability distributions for each of the parameters. 
Assuming there are $m$ parameters under investigation, the posterior will be an $m$-dimensional probability distribution. 
The result of a Bayesian sampling process is a ``chain'' consisting of $n$ samples for each parameter. 
Therefore, the full chain has a shape $(m, n)$. 
Typically these are histogrammed to show the probability of different values of the parameters. 
However, to identify independent (non-correlated) samples in the chain, autocorrelation analysis~\cite{sokal_monte_1997} may be performed and the chain ``thinned''. 
We will not cover in detail autocorrelation analysis other than to say that it helps to identify the length of separation required for samples to be independent and that thinning means that we only included samples separated by this length in the final chain.
Additionally, it is valuable to report the use of convergence diagnostics, such as the Gelman-Rubin statistic~\cite{gelman_inference_1992}, which can assist in determining if a chain appropriately describes a posterior. 

Either the full posterior chain or the thinned chain should be shared along with details of any autocorrelation analysis to accompany any Bayesian or sampling analysis. 
This will allow the best replication and verification of any results obtained from the data. 
Furthermore, large output files such as these chains can be easily shared using some versioned data repository, such as Zenodo~\cite{european_zenodo_2013} or those available at specific institutions. 
Additionally, to allow the reader to quickly interpret the sampled posterior, a graphical description (such as that in Figure~\ref{fig:post}) should be included, at a minimum, in the supplementary information of the work. 
The importance of presenting the full posterior graphically lies in the ability to easily interpret correlations between parameters through this medium.
For example, in Figure 3, the ellipsoidal probability distribution (for the $d$/$\rho$) parameters indicates correlation.
\begin{figure}
  \script{post.py}
  \includegraphics[width=\columnwidth]{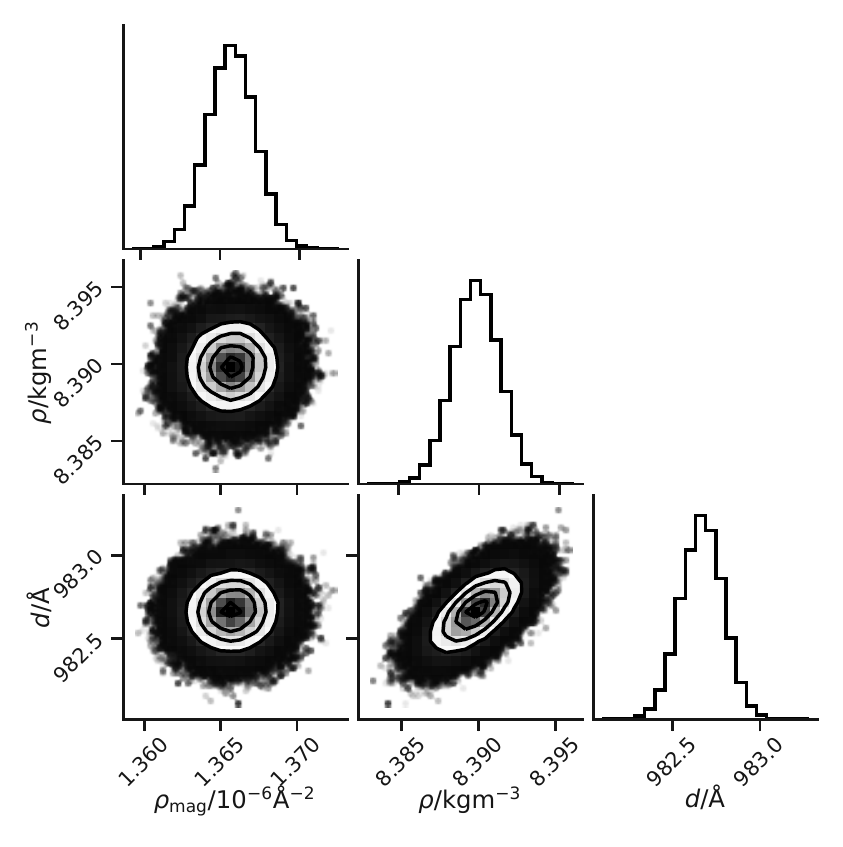}
  \caption{
    An example of a graphical depiction of the unthinned posterior as a corner plot (produced using the \texttt{corner.py} package~\cite{foremanmackey_corner_2016}), representing a three-dimensional probability distribution showing the posterior distribution for the parameters of nickel magnetic scattering length density, nickel mass density, and nickel layer thickness, from the analysis of a nickel layer on a silicon block~\cite{caruana_zenodo_2022}.
  }\label{fig:post}
\end{figure}

To report values for parameters and some form of statistical uncertainty, two approaches can be taken from the posterior chain. 
The first is to use some known statistical distribution that describes the samples well.
This is best defined for a normal distribution, for which there are statistical tests to check normality, such as the D'Agostino and Pearson's test~\cite{dagostino_omnibus_1971,dagostino_tests_1973} (which is available in the SciPy library as \texttt{scipy.stats.normaltest}~\cite{virtanen_scipy_2020}).
As with all statistical tests, this requires some threshold value to be defined to reject the null hypothesis, for this value we recommend \num{0.001} but accept that this is at the discretion of the user. 
If the parameter distribution passes a statistical test for a given distribution type, this can be quoted in the work, with information about the distribution type and the threshold value used, and the distribution can be described based on fitted parameters of the distribution as is discussed above for the Gaussian distribution. 
For example, the three parameters in Figure~\ref{fig:post} pass this statistical tests, with $p$-values of greater than \num{0.01} when \num{1000} random samples are used, therefore we can quote the parameters as normal distributions; %
  $\rho_{\mathrm{mag}}=\SI{1.366 \pm 0.001}{10^{-6}\angstrom^{-2}}$\label{output/mag_sld.txt}\unskip%
, %
  $\rho_m=\SI{8.390 \pm 0.001}{\kilogram\meter^{-3}}$\label{output/density.txt}\unskip%
, and %
  $d=\SI{982.668 \pm 0.121}{\angstrom}$\label{output/thick.txt}\unskip%
.

If it is not possible to describe the $m$-dimensional distribution using some statistical test and a common distribution type, then confidence intervals can be given. 
Where these are used the percentage of the confidence interval must be defined alongside it. 
Alongside these confidence intervals, it is typically most accurate to give the maximum probability value for the parameter, rather than the numerical mean which may sit in a region of low probability.
When reporting these quantiles of interest, we should assess how much precision we prescribe to them, which is typically achieved by defining some Monte Carlo standard error (MCSE)~\cite{vehtari_rank_2021}. 
This is the variability that would be observed should the sampling process be repeated. 
There is a range of approaches to compute the MCSE, including the \texttt{mcse} method from the \texttt{ArviZ} package~\cite{kumar_arviz_2019}.
It is important to check that the MCSE is small enough to report the level of precision desired for a given parameter. 

Regardless of how the chain is shared, as a component of a fully reproducible analysis, the author should give details of the software packages, scripts, and data used to produce the analysis and any random number seeds that were defined. 
This means that if the chain is not available, the reader can rerun the sampling and replicate the results. 
Included in this is information regarding specific version numbers for different software packages, as these can create irreplicable results between version numbers. 
We want to emphasize the value of openly sharing the posteriors of some Bayesian sampling approaches. 
All posteriors may be utilised as prior probabilities in subsequent analyses, therefore by sharing posteriors we enable improved analysis in future. 

\section{Conclusion}\label{sec:conclusions}

The use of Bayesian analysis in neutron and X-ray reflectometry is increasing, and alongside this, there is a need for analytical clarity and reproducibility. 
We have outlined the best practice, based on experience, for reporting information from Bayesian analysis. 
Specifically, we have outlined how the prior probabilities used to inform our analyses should be stated, either as uniform or more informed probability distributions that may or maybe not be described mathematically. 
We mentioned the importance of including the specific likelihood function used in an analysis. 
Additionally, we described how best to present the results from our Bayesian analysis in a clear and precise fashion, including the importance of statistical tests and confidence intervals in reporting.
We hope that this advice will be taken on by the reflectometry community and in future, there will be greater consistency and clarity in the reporting of results from Bayesian methods.
Furthermore, we hope that developers of analysis software will take this work as a call to arms to include these best practices as easy-to-access methods in their software.
Finally, if the results of neutron and X-ray Bayesian analysis are shared as outlined in this work, then the analysis will be both reproducible and comprehensible.

\section*{Data availability}

Electronic Supplementary Information (ESI) available: All analysis/plotting scripts and data files allowing for a fully reproducible and automated analysis workflow, using \showyourwork~\cite{luger_showyourwork_2022,luger_mapping_2011}, for this work is available at \url{https://github.com/arm61/reporting_sampling} (DOI:\ 10.5281/zenodo.6874559) under an MIT license, while the paper is shared under a CC BY-SA 4.0 license~\cite{mccluskey_github_2022} and the data shown in Figure~\ref{fig:post} is also available under a CC BY-SA 4.0 license~\cite{caruana_zenodo_2022}.

\section*{CR\lowercase{e}d\lowercase{i}T author statement}\label{sec:credit}

A.R.M:\ Conceptualization, Methodology, Resources, Writing --- original draft, Writing --- review \& editing, Visualisation, Project administration.
A.J.C \& C.J.K.:\ Conceptualization, Writing --- review \& editing.
Other authors:\ Writing --- review \& editing.

\begin{acknowledgments}
    All authors gratefully acknowledge the contributions of Rev. T. Bayes (\numrange{1701}{1761}) and D. S. Sivia, an early pioneer in using Bayesian methods in reflectometry analysis.
\end{acknowledgments}

\bibliographystyle{naturemag}
\bibliography{bib}

\begin{thebibliography}{10}
\expandafter\ifx\csname url\endcsname\relax
  \def\url#1{\texttt{#1}}\fi
\expandafter\ifx\csname urlprefix\endcsname\relax\def\urlprefix{URL }\fi
\providecommand{\bibinfo}[2]{#2}
\providecommand{\eprint}[2][]{\url{#2}}

\bibitem{lovell_analysis_1999}
\bibinfo{author}{Lovell, M.~R.} \& \bibinfo{author}{Richardson, R.~M.}
\newblock \bibinfo{title}{Analysis methods in neutron and {X}-ray
  reflectometry}.
\newblock \emph{\bibinfo{journal}{Curr. Opin. Colloid Interface Sci.}}
  \textbf{\bibinfo{volume}{4}}, \bibinfo{pages}{197--204}
  (\bibinfo{year}{1999}).

\bibitem{majkrzak_exact_1995}
\bibinfo{author}{Majkrzak, C.~F.} \& \bibinfo{author}{Berk, N.~F.}
\newblock \bibinfo{title}{{E}xact determination of the phase in neutron
  reflectometry}.
\newblock \emph{\bibinfo{journal}{Phys. Rev. B}} \textbf{\bibinfo{volume}{52}},
  \bibinfo{pages}{10827--10830} (\bibinfo{year}{1995}).

\bibitem{sivia_analysis_1991}
\bibinfo{author}{Sivia, D.}, \bibinfo{author}{Hamilton, W.} \&
  \bibinfo{author}{Smith, G.}
\newblock \bibinfo{title}{{A}nalysis of neutron reflectivity data: maximum
  entropy, {B}ayesian spectral analysis and speckle holography}.
\newblock \emph{\bibinfo{journal}{Physica B: Condens. Matter}}
  \textbf{\bibinfo{volume}{173}}, \bibinfo{pages}{121--138}
  (\bibinfo{year}{1991}).

\bibitem{geoghegan_experimental_1996}
\bibinfo{author}{Geoghegan, M.}, \bibinfo{author}{Jones, R. A.~L.},
  \bibinfo{author}{Sivia, D.~S.}, \bibinfo{author}{Penfold, J.} \&
  \bibinfo{author}{Clough, A.~S.}
\newblock \bibinfo{title}{{E}xperimental study of surface segregation and
  wetting in films of a partially miscible polymer blend}.
\newblock \emph{\bibinfo{journal}{Phys. Rev. E}} \textbf{\bibinfo{volume}{53}},
  \bibinfo{pages}{825--837} (\bibinfo{year}{1996}).

\bibitem{sivia_bayesian_1998}
\bibinfo{author}{Sivia, D.} \& \bibinfo{author}{Webster, J.}
\newblock \bibinfo{title}{{T}he {B}ayesian approach to reflectivity data}.
\newblock \emph{\bibinfo{journal}{Physica B: Condens. Matter}}
  \textbf{\bibinfo{volume}{248}}, \bibinfo{pages}{327--337}
  (\bibinfo{year}{1998}).

\bibitem{kienzle_refl1d_2021}
\bibinfo{author}{Kienzle, P.~A.}, \bibinfo{author}{Krycka, J.},
  \bibinfo{author}{Patel, N.} \& \bibinfo{author}{Sahin, I.}
\newblock \bibinfo{title}{Refl1d 0.8.15}.
\newblock \bibinfo{howpublished}{\url{https://refl1d.readthedocs.io/}}
  (\bibinfo{year}{2021}).

\bibitem{nelson_refnx_2019}
\bibinfo{author}{Nelson, A. R.~J.} \& \bibinfo{author}{Prescott, S.~W.}
\newblock \bibinfo{title}{\textit{refnx}: neutron and {X}-ray reflectometry
  analysis in {P}ython}.
\newblock \emph{\bibinfo{journal}{J. Appl. Crystallogr.}}
  \textbf{\bibinfo{volume}{52}}, \bibinfo{pages}{193--200}
  (\bibinfo{year}{2019}).

\bibitem{koutsioubas_anaklasis_2021}
\bibinfo{author}{Koutsioubas, A.}
\newblock \bibinfo{title}{anaklasis: a compact software package for model-based
  analysis of specular neutron and {X}-ray reflectometry data sets}.
\newblock \emph{\bibinfo{journal}{J. Appl. Crystallogr.}}
  \textbf{\bibinfo{volume}{54}}, \bibinfo{pages}{1857--1866}
  (\bibinfo{year}{2021}).

\bibitem{hughes_rascal_2019}
\bibinfo{author}{Hughes, A.~V.}
\newblock \bibinfo{title}{Rascal v1.1.0}.
\newblock
  \bibinfo{howpublished}{\url{https://github.com/arwelHughes/RasCAL_2019}}
  (\bibinfo{year}{2021}).

\bibitem{kienzle_bumps_2021}
\bibinfo{author}{Kienzle, P.~A.}, \bibinfo{author}{Krycka, J.},
  \bibinfo{author}{Patel, N.} \& \bibinfo{author}{Sahin, I.}
\newblock \bibinfo{title}{Bumps 0.8.1}.
\newblock \bibinfo{howpublished}{\url{https://bumps.readthedocs.io/}}
  (\bibinfo{year}{2021}).

\bibitem{foremanmackey_emcee_2019}
\bibinfo{author}{Foreman-Mackey, D.} \emph{et~al.}
\newblock \bibinfo{title}{emcee v3: {A} {P}ython ensemble sampling toolkit for
  affine-invariant {MCMC}}.
\newblock \emph{\bibinfo{journal}{J. Open Source Softw.}}
  \textbf{\bibinfo{volume}{4}}, \bibinfo{pages}{1864} (\bibinfo{year}{2019}).

\bibitem{speagle_dynesty_2020}
\bibinfo{author}{Speagle, J.~S.}
\newblock \bibinfo{title}{dynesty: a dynamic nested sampling package for
  estimating {B}ayesian posteriors and evidences}.
\newblock \emph{\bibinfo{journal}{Mon. Notices Royal Astron. Soc.}}
  \textbf{\bibinfo{volume}{493}}, \bibinfo{pages}{3132--3158}
  (\bibinfo{year}{2020}).

\bibitem{mccluskey_bayesian_2019}
\bibinfo{author}{McCluskey, A.~R.} \emph{et~al.}
\newblock \bibinfo{title}{{B}ayesian determination of the effect of a deep
  eutectic solvent on the structure of lipid monolayers}.
\newblock \emph{\bibinfo{journal}{Phys. Chem. Chem. Phys.}}
  \textbf{\bibinfo{volume}{21}}, \bibinfo{pages}{6133--6141}
  (\bibinfo{year}{2019}).

\bibitem{mccluskey_general_2020}
\bibinfo{author}{McCluskey, A.~R.}, \bibinfo{author}{Cooper, J. F.~K.},
  \bibinfo{author}{Arnold, T.} \& \bibinfo{author}{Snow, T.}
\newblock \bibinfo{title}{{A} general approach to maximise information density
  in neutron reflectometry analysis}.
\newblock \emph{\bibinfo{journal}{Mach. Learn.: Sci. Technol.}}
  \textbf{\bibinfo{volume}{1}}, \bibinfo{pages}{035002} (\bibinfo{year}{2020}).

\bibitem{sivia_data_2006}
\bibinfo{author}{Sivia, D.~S.} \& \bibinfo{author}{Skelling, J.}
\newblock \emph{\bibinfo{title}{{D}ata {A}nalysis : a {B}ayesian tutorial}}
  (\bibinfo{publisher}{Oxford University Press}, \bibinfo{address}{Oxford, GB},
  \bibinfo{year}{2006}).

\bibitem{hughes_physical_2019}
\bibinfo{author}{Hughes, A.~V.} \emph{et~al.}
\newblock \bibinfo{title}{{P}hysical {P}roperties of {B}acterial {O}uter
  {M}embrane {M}odels: {N}eutron {R}eflectometry \& {M}olecular {S}imulation}.
\newblock \emph{\bibinfo{journal}{Biophys. J.}} \textbf{\bibinfo{volume}{116}},
  \bibinfo{pages}{1095--1104} (\bibinfo{year}{2019}).

\bibitem{aboljadayel_determining_2021}
\bibinfo{author}{Aboljadayel, R. O.~M.} \emph{et~al.}
\newblock \bibinfo{title}{{D}etermining the {P}roximity {E}ffect {I}nduced
  {M}agnetic {M}oment in {G}raphene by {P}olarized {N}eutron {R}eflectivity and
  {X}-ray {M}agnetic {C}ircular {D}ichroism} (\bibinfo{year}{2021}).
\newblock \eprint{arXiv:2101.09946}.

\bibitem{bevington_data_2002}
\bibinfo{author}{Bevington, P.} \& \bibinfo{author}{Robinson, D.~K.}
\newblock \emph{\bibinfo{title}{{D}ata {R}eduction and {E}rror {A}nalysis for
  the {P}hysical {S}ciences}} (\bibinfo{publisher}{McGraw-Hill Education},
  \bibinfo{address}{New York City, US}, \bibinfo{year}{2002}),
  \bibinfo{edition}{3} edn.

\bibitem{knoops_atomic_2015}
\bibinfo{author}{Knoops, H. C.~M.} \emph{et~al.}
\newblock \bibinfo{title}{{A}tomic {L}ayer {D}eposition of {S}ilicon {N}itride
  from {B}is({\emph{tert}}-butylamino)silane and {N$_2$} {P}lasma}.
\newblock \emph{\bibinfo{journal}{ACS Appl. Mater. Interfaces}}
  \textbf{\bibinfo{volume}{7}}, \bibinfo{pages}{19857--19862}
  (\bibinfo{year}{2015}).

\bibitem{lass_multinomial_2021}
\bibinfo{author}{Lass, J.}, \bibinfo{author}{B{\o}ggild, M.~E.},
  \bibinfo{author}{Hedeg{\aa}rd, P.} \& \bibinfo{author}{Lefmann, K.}
\newblock \bibinfo{title}{Multinomial, poisson and gaussian statistics in count
  data analysis}.
\newblock \emph{\bibinfo{journal}{Journal of Neutron Research}}
  \textbf{\bibinfo{volume}{23}}, \bibinfo{pages}{69--92}
  (\bibinfo{year}{2021}).

\bibitem{sokal_monte_1997}
\bibinfo{author}{Sokal, A.}
\newblock \bibinfo{title}{{M}onte {C}arlo {M}ethods in {S}tatistical
  {M}echanics: {F}oundations and {N}ew {A}lgorithms}.
\newblock In \bibinfo{editor}{DeWitt-Morette, C.}, \bibinfo{editor}{Cartier,
  P.} \& \bibinfo{editor}{Folacci, A.} (eds.)
  \emph{\bibinfo{booktitle}{{F}unctional {I}ntegration: {B}asics and
  {A}pplications}}, chap.~\bibinfo{chapter}{6}, \bibinfo{pages}{131--192}
  (\bibinfo{publisher}{Springer}, \bibinfo{address}{New York City, US},
  \bibinfo{year}{1997}).

\bibitem{gelman_inference_1992}
\bibinfo{author}{Gelman, A.} \& \bibinfo{author}{Rubin, D.~B.}
\newblock \bibinfo{title}{{I}nference from {I}terative {S}imulation using
  {M}ultiple {S}equences}.
\newblock \emph{\bibinfo{journal}{Stat. Sci.}} \textbf{\bibinfo{volume}{7}},
  \bibinfo{pages}{457--472} (\bibinfo{year}{1992}).

\bibitem{european_zenodo_2013}
\bibinfo{author}{{European Organization For Nuclear Research}} \&
  \bibinfo{author}{{OpenAIRE}}.
\newblock \bibinfo{title}{Zenodo} (\bibinfo{year}{2013}).

\bibitem{foremanmackey_corner_2016}
\bibinfo{author}{Foreman-Mackey, D.}
\newblock \bibinfo{title}{corner.py: {S}catterplot matrices in {P}ython}.
\newblock \emph{\bibinfo{journal}{J. Open Source Softw.}}
  \textbf{\bibinfo{volume}{1}}, \bibinfo{pages}{24} (\bibinfo{year}{2016}).

\bibitem{caruana_zenodo_2022}
\bibinfo{author}{Caruana, A.~J.} \& \bibinfo{author}{Kinane, C.~J.}
\newblock \bibinfo{title}{Chains from {N}i58 standard sample}
  (\bibinfo{year}{2022}).

\bibitem{dagostino_omnibus_1971}
\bibinfo{author}{D'Agstino, R.~B.}
\newblock \bibinfo{title}{{A}n omnibus test of normality for moderate and large
  size samples}.
\newblock \emph{\bibinfo{journal}{Biometrika}} \textbf{\bibinfo{volume}{58}},
  \bibinfo{pages}{341--348} (\bibinfo{year}{1971}).

\bibitem{dagostino_tests_1973}
\bibinfo{author}{D'Agostino, R.} \& \bibinfo{author}{Pearson, E.~S.}
\newblock \bibinfo{title}{{T}ests for departure from normality}.
\newblock \emph{\bibinfo{journal}{Biometrika}} \textbf{\bibinfo{volume}{60}},
  \bibinfo{pages}{613--622} (\bibinfo{year}{1973}).

\bibitem{virtanen_scipy_2020}
\bibinfo{author}{Virtanen, P.} \emph{et~al.}
\newblock \bibinfo{title}{{SciPy} 1.0: fundamental algorithms for scientific
  computing in {P}ython}.
\newblock \emph{\bibinfo{journal}{Nat. Methods}} \textbf{\bibinfo{volume}{17}},
  \bibinfo{pages}{261--272} (\bibinfo{year}{2020}).

\bibitem{vehtari_rank_2021}
\bibinfo{author}{Vehtari, A.}, \bibinfo{author}{Gelman, A.},
  \bibinfo{author}{Simpson, D.}, \bibinfo{author}{Carpenter, B.} \&
  \bibinfo{author}{Bü\"{u}kner, P.-C.}
\newblock \bibinfo{title}{{R}ank-{N}ormalization, {F}olding, and
  {L}ocalization: {A}n {I}mproved $\widehat{R}$ for {A}ssessing {C}onvergence
  of {MCMC} (with {D}iscussion)}.
\newblock \emph{\bibinfo{journal}{Bayesian Anal.}}
  \textbf{\bibinfo{volume}{16}}, \bibinfo{pages}{667--718}
  (\bibinfo{year}{2021}).

\bibitem{kumar_arviz_2019}
\bibinfo{author}{Kumar, R.}, \bibinfo{author}{Carroll, C.},
  \bibinfo{author}{Hartikainen, A.} \& \bibinfo{author}{Martin, O.}
\newblock \bibinfo{title}{{A}rvi{Z} a unified library for exploratory analysis
  of {B}ayesian models in python}.
\newblock \emph{\bibinfo{journal}{J. Open Source Softw.}}
  \textbf{\bibinfo{volume}{4}}, \bibinfo{pages}{1143} (\bibinfo{year}{2019}).

\bibitem{luger_showyourwork_2022}
\bibinfo{author}{{Luger}, R.}
\newblock \bibinfo{title}{{showyourwork}}.
\newblock
  \bibinfo{howpublished}{\url{https://github.com/showyourwork/showyourwork}}
  (\bibinfo{year}{2022}).

\bibitem{luger_mapping_2011}
\bibinfo{author}{Luger, R.} \emph{et~al.}
\newblock \bibinfo{title}{{M}apping stellar surfaces {III}: {A}n {E}fficient,
  {S}calable, and {O}pen-{S}ource {D}oppler {I}maging {M}odel}
  (\bibinfo{year}{2021}).
\newblock \eprint{arXiv:2110.06271}.

\bibitem{mccluskey_github_2022}
\bibinfo{author}{McCluskey, A.~R.} \emph{et~al.}
\newblock \bibinfo{title}{{ESI for ``Advice on describing Bayesian analysis of
  neutron and X-ray reflectometry''}} (\bibinfo{year}{2022}).
\newblock \urlprefix\url{https://github.com/arm61/reporting_sampling}.

\end{thebibliography}

\end{document}